\documentclass[lettersize,journal]{IEEEtran}
\usepackage{amsmath}
\usepackage{amsmath,amsfonts}
\usepackage{algorithmic}
\usepackage{algorithm}
\usepackage{array}
\usepackage[caption=false,font=footnotesize,labelfont=rm,textfont=rm]{subfig}
\usepackage{textcomp}
\usepackage{stfloats}
\usepackage{url}
\usepackage{verbatim}
\usepackage{graphicx}
\usepackage{cite}
\usepackage[T1]{fontenc}
\usepackage{booktabs,multirow}
\usepackage{CJKutf8}
\usepackage{pifont}
\usepackage{color}
\usepackage{xcolor}
\usepackage{hyperref} 
\hypersetup{hidelinks,
  colorlinks=ture,
  linkcolor=black
  }

\begin{document}
\begin{CJK}{UTF8}{gbsn}

\title{Interference Analysis for Coexistence of UAVs and Civil Aircrafts Based on Automatic Dependent Surveillance-Broadcast}

\author{Yiyang Liao, Ziye Jia, \IEEEmembership{Member,~IEEE,} Chao Dong, \IEEEmembership{Member,~IEEE,} Lei Zhang, \\
Qihui Wu, \IEEEmembership{Fellow,~IEEE}, Huiling Hu, and Zhu Han, \IEEEmembership{Fellow,~IEEE}
\thanks{Yiyang Liao, Chao Dong, Lei Zhang and Qihui Wu are with the College of Electronic and
Information Engineering, Nanjing University of Aeronautics and Astronautics,
Nanjing 211106, China (e-mail: liaoyiyang@nuaa.edu.cn; dch@nuaa.edu.cn; Zhang\_lei@nuaa.edu.cn; wuqihui@nuaa.edu.cn).}
\thanks{ Ziye Jia is with the College of Electronic and Information Engineering, Nanjing University of Aeronautics and Astronautics, Nanjing 211106, China, and also with the National Mobile Communications Research Laboratory, Southeast University, Nanjing, Jiangsu, 211111, China (e-mail: jiaziye@nuaa.edu.cn).}
\thanks{Huiling Hu is with the Middle-south Regional Air Traffic Management Bureau of CAAC (e-mail: hhl@atmb.org).}
\thanks{Zhu Han is with the Department of Electrical and Computer Engineering,
University of Houston, Houston, TX 77004 USA, and also with the
Department of Computer Science and Engineering, Kyung Hee University,
Seoul 446-701, South Korea (e-mail: hanzhu22@gmail.com).}
\thanks{\emph{Corresponding author: Ziye Jia and Chao Dong. }}
}
  


\maketitle
\pagestyle{empty}  
\thispagestyle{empty}
\begin{abstract}
Due to the advantages of high mobility and easy deployment, unmanned aerial vehicles (UAVs) are widely applied in both military and civilian fields. In order to strengthen the flight surveillance of UAVs and guarantee the airspace safety, UAVs can be equipped with the automatic dependent surveillance-broadcast (ADS-B) system, which periodically sends flight information to other aircrafts and ground stations (GSs). However, due to the limited resource of channel capacity, UAVs equipped with ADS-B results in the interference between UAVs and civil aircrafts (CAs), which further impacts the accuracy of received information at GSs. In detail, the channel capacity is mainly affected by the density of aircrafts and the transmitting power of ADS-B. Hence, based on the three-dimensional poisson point process, this work leverages the stochastic geometry theory to build a model of the coexistence of UAVs and CAs and analyze the  interference performance of ADS-B monitoring system. From simulation results, we reveal the effects of transmitting power, density, threshold and pathloss on the performance of the ADS-B monitoring system. Besides, we provide the suggested transmitting power and density for the safe coexistence of UAVs and CAs.
\end{abstract}


\section{Introduction}\label{s1}
\IEEEPARstart{A}{s} low-altitude aerial technology advances, unmanned aerial vehicles (UAVs) play increasingly significant roles across various domains such as collaborative reconnaissance, precision agriculture, disaster rescue, and environmental monitoring \cite{ref1}. Besides, the growing multitude applications necessitate abundant UAVs, raising concerns about the flight safety. Furthermore, the absence of on-board pilots poses potential risks to the safe operation of civil aircrafts (CAs) \cite{ref2}. Hence, enhancing the airspace management for UAVs and guaranteeing the flight safety emerge as crucial imperatives \cite{ref3}.
\par In order to obtain exact aerial location information, UAVs can be equipped with automatic dependent surveillance-broadcast (ADS-B) systems\cite{ref4}. Working at 1090MHz, ADS-B is beneficial to both the flight safety and air traffic management \cite{ref5}. In detail, an aircraft equipped with ADS-B can automatically broadcast its flight information to nearby aircrafts and ground stations (GSs)\cite{ref6}. However, due to the limited channel capacity, if multiple UAVs utilize the same channel, the interference between UAVs and CAs cannot be neglected. Consequently, the accuracy of received information  at GSs is deteriorated, which further impairs the performance of the  monitoring system. In particular, the density of UAVs and the transmitting power of ADS-B are key factors on the performance of the monitoring system. As a result, this work aims to analyze the interference for the coexistence of UAVs and CAs based on ADS-B with respect to the transmitting power and the density of UAVs.
\IEEEpubidadjcol
\par There exist a couple of related works about UAVs equipped with ADS-B. For instance, \cite{ref7} points out that the performance of ADS-B system is affected by the density of UAVs, ADS-B transmitting power and the number of GSs. \cite{ref8} demonstrates that the performance of the ADS-B system is affected by channel characteristics and minimum updating interval required by aircrafts. Besides, there are a few works utilizing stochastic geometry (SG) and poisson point process (PPP) to analyze UAVs networks, for example, \cite{ref9} states that UAV wireless networks have natural spatial random characteristics and the channel has fading and shadowing characteristics. Therefore, SG can be utilized to analyze the performance of UAV wireless networks. \cite{ref10} develops a tractable framework for signal-to-interference-plus-noise ratio (SINR) analysis in downlink heterogeneous cellular networks with flexible cell association policies. \cite{ref11} compares the model of Rician channel, Rayleigh channel and Nakagami-m channel in wireless network, and analyzes the coverage rate of UAV assisted cellular network. The authors in \cite{ref12} utilize SG to analyze the response delay and the successful transmission probability for a single link and a group of links based on the three-dimensional (3D) distribution of UAV swarms. 
\par However, the above works equip UAVs with ADS-B to enhance the flight safety, but they seldom utilize SG to examine the system performance considering the interference based on ADS-B. Besides, to construct the model, we distinguish the interference taking account of the coexistence of UAVs and CAs, which is a challenging problem. In, short, the contributions of this paper are summarized as follows. 
\begin{itemize}
\item We propose the model of the coexistence of UAVs and CAs based on ADS-B techniques.
\item Based on SG theory, we employ 3D-PPP to reveal the interference performance and deduce the analytic form of the received probability.
\item Extensive simulations are conducted to verify the effects of transmitting
power, density, threshold and pathloss on the performance of ADS-B monitoring system.
\end{itemize}
\section{System Model}\label{s2}

\subsection{Network Model}
\begin{figure}[t]
  \centerline{\includegraphics[width=0.8\linewidth]{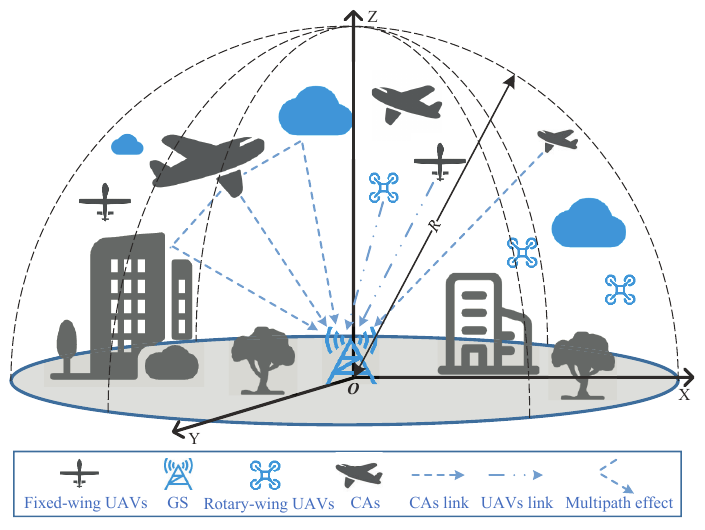}}
  \caption{System model of the coexistence of UAVs and CAs.}
  \label{f1}
\end{figure}
Fig. \ref{f1} depicts the system model of the coexistence of UAVs and CAs. The fixed-wing UAVs, rotary-wing UAVs and CAs are randomly distributed in space \textsl{\textbf{V}}. All aircrafts transmit flight information to the GS via ADS-B. The fixed-wing UAVs and the rotary-wing UAVs follow a 3D-PPP \cite{ref13} with density $λ\lambda_1$ in the finite space \textsl{\textbf{V}}, and the number of UAVs is $N_U=λ\lambda_1\textsl{\textbf{V}}$. The CAs follow a 3D-PPP with density $λ\lambda_2$ in the finite space  \textsl{\textbf{V}}, and the number of CAs is $N_C=λ\lambda_2\textsl{\textbf{V}}$. Denote the UAV set as $\mathcal U=\{U_1,...U_i,...,U_{N_U}\}$, and the CA set as $\mathcal C=\{C_1,...C_j,...,C_{N_C}\}$. It is assumed that there is only one GS in space \textsl{\textbf{V}}, and all ADS-B packets from UAVs and CAs are received by the GS. The GS is located in the center of the ground with the coordinate of \textsl{\textbf{O}}(0, 0, 0). In detail, the coordinate of the $i$-$th$ UAV in  set $\mathcal U$ is $(x_{U_i}$, $y_{U_i}$, $z_{U_i})$ and the coordinate of the $j$-$th$ CA in set $\mathcal C$ is $(x_{C_j}$, $y_{C_j}$, $z_{C_j})$. The X-axis coordinates for all aircrafts range within $[-L_x, L_x]$, the Y-axis coordinates range within $[-L_y, L_y]$, and the Z-axis coordinates are $[0, L_z]$. The euclidean distance between UAV $U_i$ and the GS is $d_{U_i}=\sqrt{{x^2_{U_i}}+{y^2_{U_i}}+{z^2_{U_i}}}$, and the euclidean distance between CA $C_j$ and the GS is $d_{C_j}=\sqrt{{x^2_{C_j}}+{y^2_{C_j}}+{z^2_{C_j}}}$. The transmitting power of ADS-B from UAV $U_i$ and CA $C_j$ are set as $P_U$ and $P_C$, respectively.
\subsection{Channel Model}
$G_{U_t}$ and $G_{C_t}$ respectively represent the transmitter gain of UAVs and CAs. $G_r$ denotes the receiver gain at the GS. Therefore, the total air-ground (AG) channel gain at the GS from UAVs is $G_U=G_{U_t}G_r$, and the AG channel gain between CAs and the GS is $G_C=G_{C_t}G_r$. The pathloss from the GS to UAVs and CAs are respectively proportional to $d_{U_i}^{-\alpha}$ and $d_{C_j}^{-\alpha}$, where $d_{U_i}$ and  $d_{C_j}$ represent the distance between the aircrafts and GS. $\alpha$ indicates the pathloss index. ${h_{U_i}}$ represents the gain of small scale fading channel between UAV $U_i$ and the GS. ${h_{C_j}}$ represents the gain of small scale fading channel between CA $C_j$ and the GS. ${h_{U_i}}$ and ${h_{C_j}}$ are two random variables following an exponential distribution with mean value of 1. Gaussian white noise $N$ is added to the model, i.e., $N=n_0\times B$, where $n_0$ is noise power density and $B$ is the system bandwidth. We leverage $\gamma$ to represent SINR. Then, the $\gamma_U^m$ of the desired signal sent by the $m$-$th$ UAV $U_m$ in set $\mathcal U$ is

\begin{equation} 
  \gamma_U^m=\frac{{P_U}{G_U}{h_{U_m}}{d_{U_m}^{-\alpha}}}{N+{P_U}I_{\mathcal {U}\backslash \{{U_m}\}}+{P_C}{I_{\mathcal{C}}}},
\end{equation}

\noindent in which 

\begin{equation} 
I_{\mathcal{U}\backslash \{{U_m}\}}=\sum_{{U_i}\in \mathcal{U}\backslash \{{U_m}\}}{G_U}{h_{U_i}}{d_{U_i}^{-\alpha}},
\end{equation}

\noindent and

\begin{equation} 
I_{\mathcal{C}}=\sum_{{C_j}\in \mathcal{C}}{G_C}{h_{C_j}}{d_{C_j}^{-\alpha}}.
\end{equation}

\section{Performance Analysis}\label{s3}
\par It is supposed that all the aircrafts send the flight information via ADS-B to the GS within space \textsl{\textbf{V}}. In particular, UAVs follow the nearest neighbor association strategy\cite{ref14}, i.e., no other GSs outside space \textsl{\textbf{V}} has less distance to the target UAV
\vspace{-0.2cm} 
\begin{equation} 
  P\{d_{U_i}\textgreater R\}={\rm exp}(-λ\lambda_1\textsl{\textbf{V}})={\rm exp}\bigg(-\frac{4}{3}\pi\lambda_1{d^3_{U_i}}\bigg),
\end{equation}

\noindent where $d_{U_i}$$\geq$0, and $R$ is the radius of the 3D euclidean space $\mathbb{R}^3$. Therefore, the cumulative distribution function (CDF) of the distance $d_{U_i}$ from UAV $U_i$ to the GS is
\vspace{-0.2cm} 
\begin{equation} 
  F_U(d_{U_i})= P\{d_{U_i}\leq R\}=1-{\rm exp}\bigg(-\frac{4}{3}\pi\lambda_1{d^3_{U_i}}\bigg),
\end{equation}

\noindent and the probability density function (PDF) of $d_{U_i}$ is
\begin{equation} 
  f_U(d_{U_i})=\frac{dF_U(d_{U_i})}{d(d_{U_i})}=4\pi\lambda_1{d^2_{U_i}}{\rm exp}\bigg(-\frac{4}{3}\pi\lambda_1{d^3_{U_i}}\bigg).
\end{equation}

The successful received probability $P_{suc}$ at the GS is introduced to measure the transmission quality. If the distance between the UAV and GS is $d$, and $\gamma$ is greater than the received threshold $\theta$, the successful received probability of the GS is denoted as
\begin{equation} 
  P_{suc}=\mathbb{E}[P(\gamma\geq\theta|d)].
\end{equation}

\noindent Since $\gamma$ is also a function of $d$, the $P_{suc}$ of the $m$-$th$ UAV in set $\mathcal U$
is further expressed as

\begin{align} \label{e8}
  P_{suc}&=\int_{0}^{\infty}P(\gamma_U^i\geq\theta|d_{U_m})f_U(d_{U_m}) \,d(d_{U_m})\nonumber\\
  &=\int_{0}^{\infty}4\pi\lambda_1{d^2_{U_m}}P(\gamma_U^i\geq\theta|d_{U_m})\nonumber\\
  &\quad  {\rm exp}\bigg(-\frac{4}{3}\pi\lambda_1{d^3_{U_m}}\bigg) \,d(d_{U_m}).
\end{align}

\noindent It is assumed that the average gain of small scale fading channel is a random variable following the Gamma distribution with mean value of 1\cite{ref13}, which is depicted as

\begin{figure}[t]
  \centerline{\includegraphics[width=0.7\linewidth]{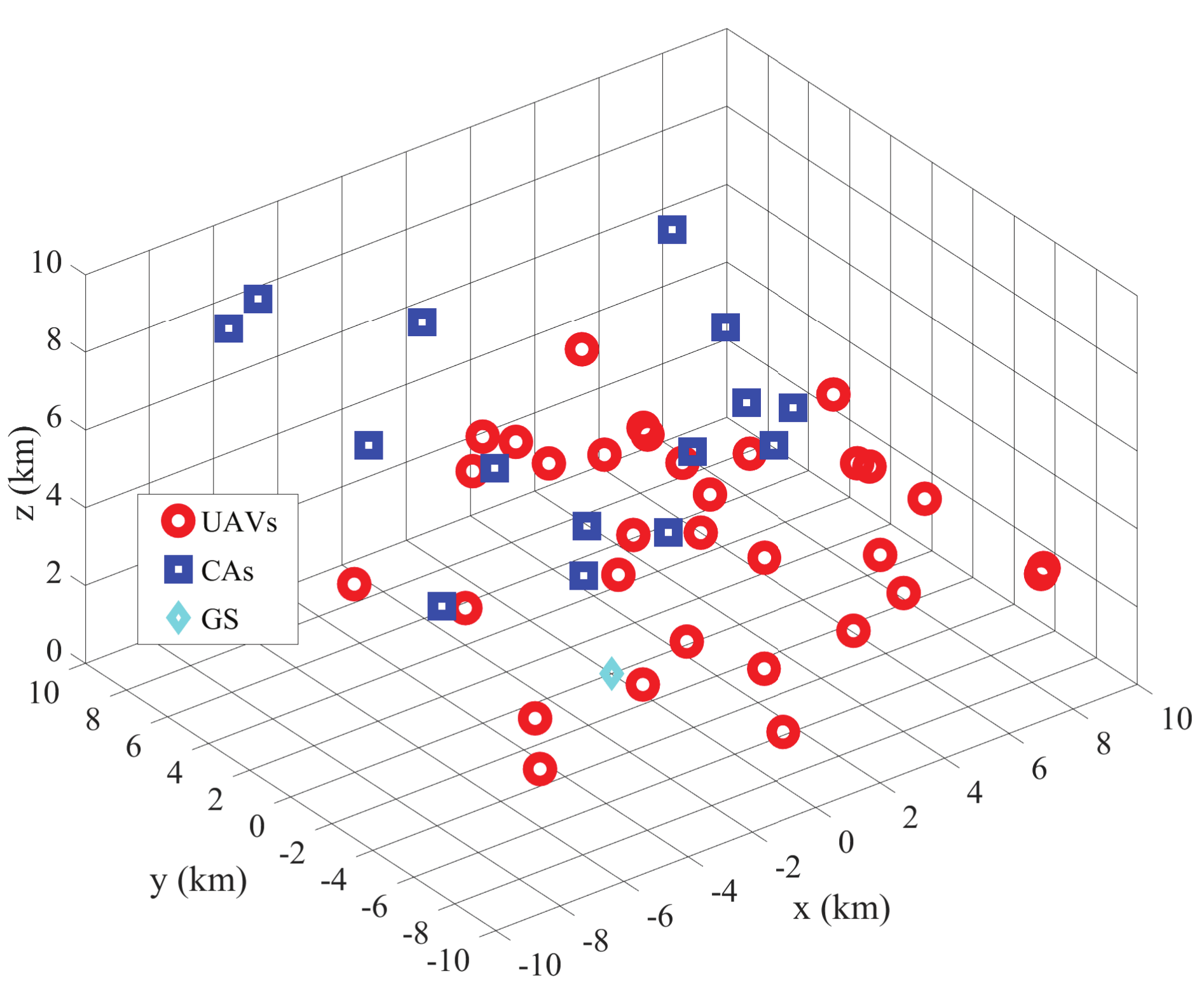}}
   \caption{Simulation scenario.}
   \label{f2}
 \end{figure}
 
 \begin{figure}[t]
   \centerline{\includegraphics[width=0.6\linewidth]{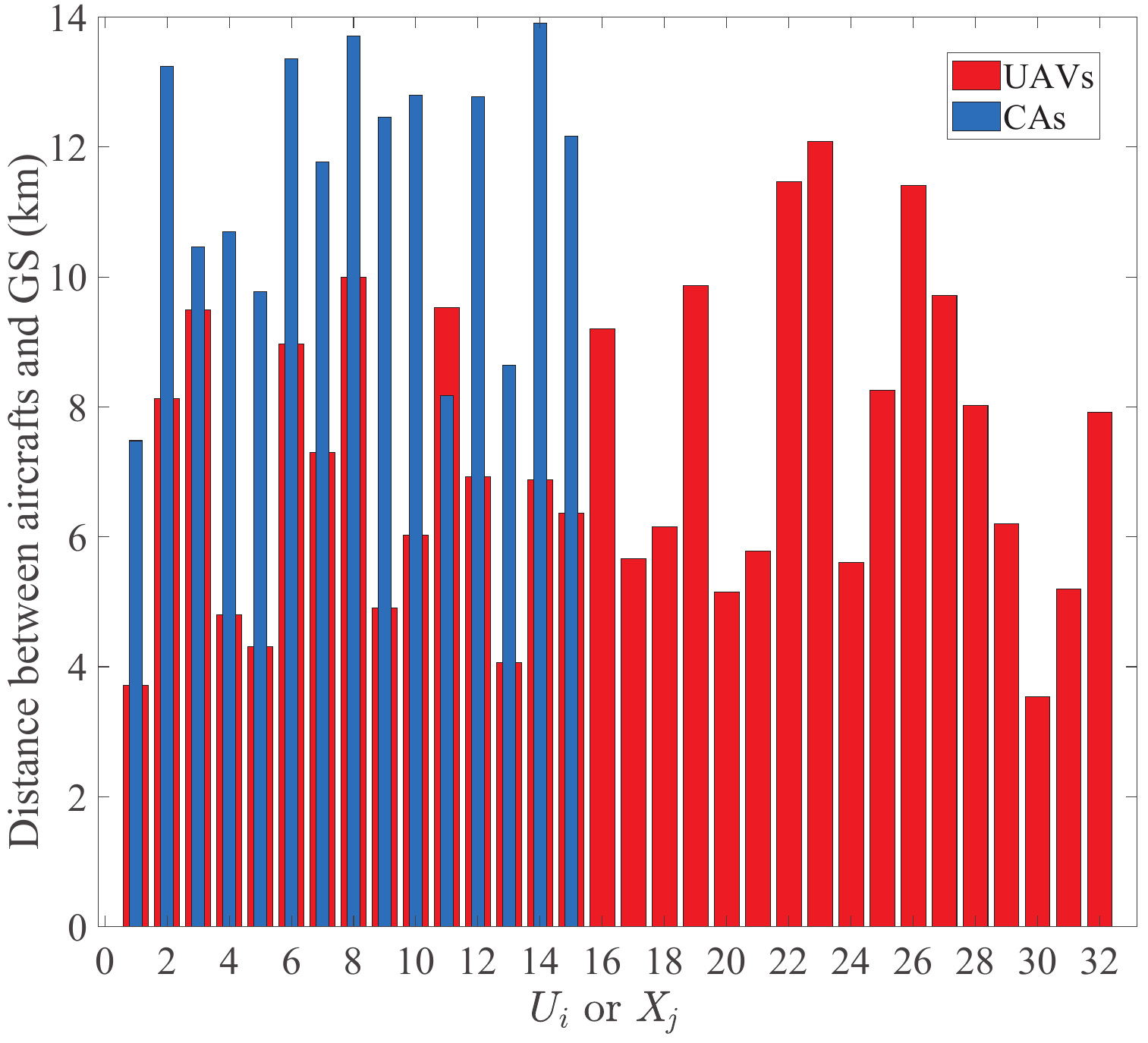}}
   \caption{Distance between GS and UAVs or CAs.}
   \label{f3}
 \end{figure}

\begin{equation} 
  f(h)=\frac{\beta^\beta  }{\Gamma(\beta )} h^{\beta-1}e^{-\beta h}.
\end{equation}

\noindent When $\beta$ equals 1, the channel is considered as Rayleigh fading. $h$ follows an exponential distribution with mean value of 1. The PDF of $h$ is $f(x)$=$e^{-x}$, i.e., ${h_{U_i}}$$\sim$${\rm exp}(1)$ and ${h_{C_j}}$$\sim$${\rm exp}(1)$. Hence, $P(\gamma_U^i\geq\theta|d_{U_m})$ in (\ref{e8}) is further represented as 
\vspace{-0.1cm} 
\begin{align}\label{e10}
  &P(\gamma_U^i\geq\theta|d_{U_m})\nonumber\\
  &=P\bigg({h_{U_m}}\geq\frac{\theta{{d^\alpha_{U_m}}}(N+{P_U}I_{\mathcal{U}\backslash \{{U_m}\}}+{P_C}{I_{\mathcal{C}}}) }{P_UG_U}\bigg)\nonumber\\
  &={\rm exp}\bigg(\frac{-\theta{{d^\alpha_{U_m}}}(N+{P_U}I_{\mathcal{U}\backslash \{{U_m}\}}+{P_C}{I_{\mathcal{C}}})}{P_UG_U}\bigg)\nonumber\\
  &={\rm exp}\bigg(\frac{-\theta{{d^\alpha_{U_m}}}N}{P_UG_U}\bigg)\mathbb{L}_{I_{\mathcal{U}\backslash \{{U_m}\}}}\bigg(\frac{\theta{{d^\alpha_{U_m}}}}{G_U}\bigg)\mathbb{L}_{I_{\mathcal{C}}}\bigg(\frac{\theta{{d^\alpha_{U_m}}}}{G_U}\times \frac{P_C}{P_U}\bigg).
\end{align}
\noindent Let $\frac{\theta{{d^\alpha_{U_m}}}}{G_U}=s_1$, and we have
\begin{equation}\label{e11}
\mathbb{L}_{I_{\mathcal{U}\backslash \{{U_m}\}}}\bigg(\frac{\theta{{d^\alpha_{U_m}}}}{G_U}\bigg)=\mathbb{L}_{I_{\mathcal{U}\backslash\{{U_m}\}}}(s_1)=\mathbb{E}[e^{-s_1(I_{\mathcal{U}\backslash \{{U_m}\}})}],
\end{equation}
\noindent which is the Laplace transform of $I_{\mathcal{U}\backslash \{{U_m}\}}$, and is further derived as
\vspace{-0.2cm} 
\begin{align}\label{e12}
  \mathbb{L}_{I_{\mathcal{U}\backslash\{{U_m}\}}}(s_1)&=\mathbb{E}\bigg[{\rm exp}\bigg(-s_1\sum_{{U_i}\in \mathcal{U}\backslash \{{U_m}\}}{G_U}{h_{U_i}}{d_{U_i}^{-\alpha}}\bigg)\bigg]\nonumber\\
  &\overset{(\textbf{a})}{=}\mathbb{E}\bigg[\prod_{U_i\in \mathcal{U}\backslash\{{U_m}\}}\frac{1}{1+s_1G_Ud_{U_i}^{-\alpha}}\bigg]\nonumber\\
  &\overset{(\textbf{b})}{=}{\rm exp}\bigg[-\lambda_1\int_\textsl{\textbf{V}}\bigg(1-\frac{1}{1+s_1G_Ud_{U_i}^{-\alpha}}\bigg)d(d_{U_i})\bigg]\nonumber\\
  &\overset{(\textbf{c})}{=}{\rm exp}\bigg[-\lambda_1\int_{-L_x}^{L_x}\int_{-L_y}^{L_y}\int_{0}^{L_z}1-\nonumber\\
  &\quad\quad\frac{1}{1+s_1G_Ud_{U_i}^{-\alpha}}dxdydz\bigg]\nonumber\\
  &\overset{(\textbf{d})}{=}{\rm exp}(-\lambda_1H_1),
\end{align}

\noindent where (\textbf{a}) is obtained by the moment generating function, (\textbf{b}) follows the probability generating function (PGFL) of the PPP \cite{ref15}, $d_{U_i}$ in (\textbf{c}) can be further expressed as $\sqrt{{x_{U_i}}^2+{y_{U_i}}^2+{z_{U_i}}^2}$, and $H_1$ in (\textbf{d}) represents the triple integral in (\textbf{c}). Moreover, let $\frac{\theta{{d^\alpha_{U_m}}}}{G_U}\times \frac{P_C}{P_U}=s_2$, and we have

\begin{equation}\label{e13}
\mathbb{L}_{I_{\mathcal{C}}}\bigg(\frac{\theta{{d^\alpha_{U_m}}}}{G_U}\times \frac{P_C}{P_U}\bigg)=\mathbb{L}_{I_{\mathcal{C}}}(s_2)=\mathbb{E}[e^{-s_2I_{\mathcal{C}}}],
\end{equation}

\noindent following the Laplace transformation of $I_{\mathcal{C}}$. Therefore, (\ref{e13}) is further simplified as
\vspace{-0.2cm} 
\begin{align}\label{e14}
    \mathbb{L}_{I_{\mathcal{C}}}(s_2)&=\mathbb{E}\bigg[{\rm exp}\bigg(-s_2\sum_{{C_j}\in \mathcal{C}}{G_C}{h_{C_j}}{d_{C_j}^{-\alpha}}\bigg)\bigg]\nonumber\\
    &=\mathbb{E}\bigg[\prod_{C_j\in \mathcal{C}}\frac{1}{1+s_2G_Cd_{C_j}^{-\alpha}}\bigg]\nonumber\\
    &\overset{(\textbf{e})}{=}{\rm exp}(-\lambda_2H_2),
\end{align}

\noindent where $H_2$ in (\textbf{e}) symbolizes the triple integral. 
By substituting (\ref{e10}), (\ref{e11}), (\ref{e12}), (\ref{e13}) and (\ref{e14}) into (\ref{e8}), $P_{suc}$ is calculated as 
\begin{align}\label{e15}
    P_{suc}&=\int_{0}^{\infty}4\pi\lambda_1{d^2_{U_m}}{\rm exp}\bigg(\frac{-\theta{{d^\alpha_{U_m}}}N}{P_UG_U}-\lambda_1H_1\nonumber\\
    &\quad-\lambda_2H_2-\frac{4}{3}\pi\lambda_1{d^3_{U_m}}\bigg)d(d_{U_m}). 
\end{align}
The influences of $H_1$ and $H_2$ on $P_{suc}$ have different weights, which are proportional to $\lambda_1$ and $\lambda_2$, respectively. According to (\ref{e12}), $H_2$ is mainly affected by the ratio of $P_C$ to $P_U$. Hence, following (\ref{e15}), $P_{suc}$ is mainly related with $\lambda_1$, $P_U$, $P_C$, $\theta$ and $\alpha$. Analyzing the performance on UAVs or CAs is a similar reasoning process, i.e., the SINR and successful received probability have similar analytic forms. The detailed difference between UAVs and CAs lies in the altitude, density and transmitting power of ADS-B. Hovover, in this paper, our research mainly focuses on the analysis of the performance of GS on UAVs, and CAs are regarded as a interfering factor. Hence, we only provide the deductions and simulations concentrated on UAVs.
Further, we analyze the detailed influences in Section \ref{s4}. 

\begin{figure*}[t] 
  \centering
 \begin{minipage}{1\linewidth}
\centering
  \subfloat[The impact of $P_U$ on the received probability.]{
  \includegraphics[width=0.31\linewidth]{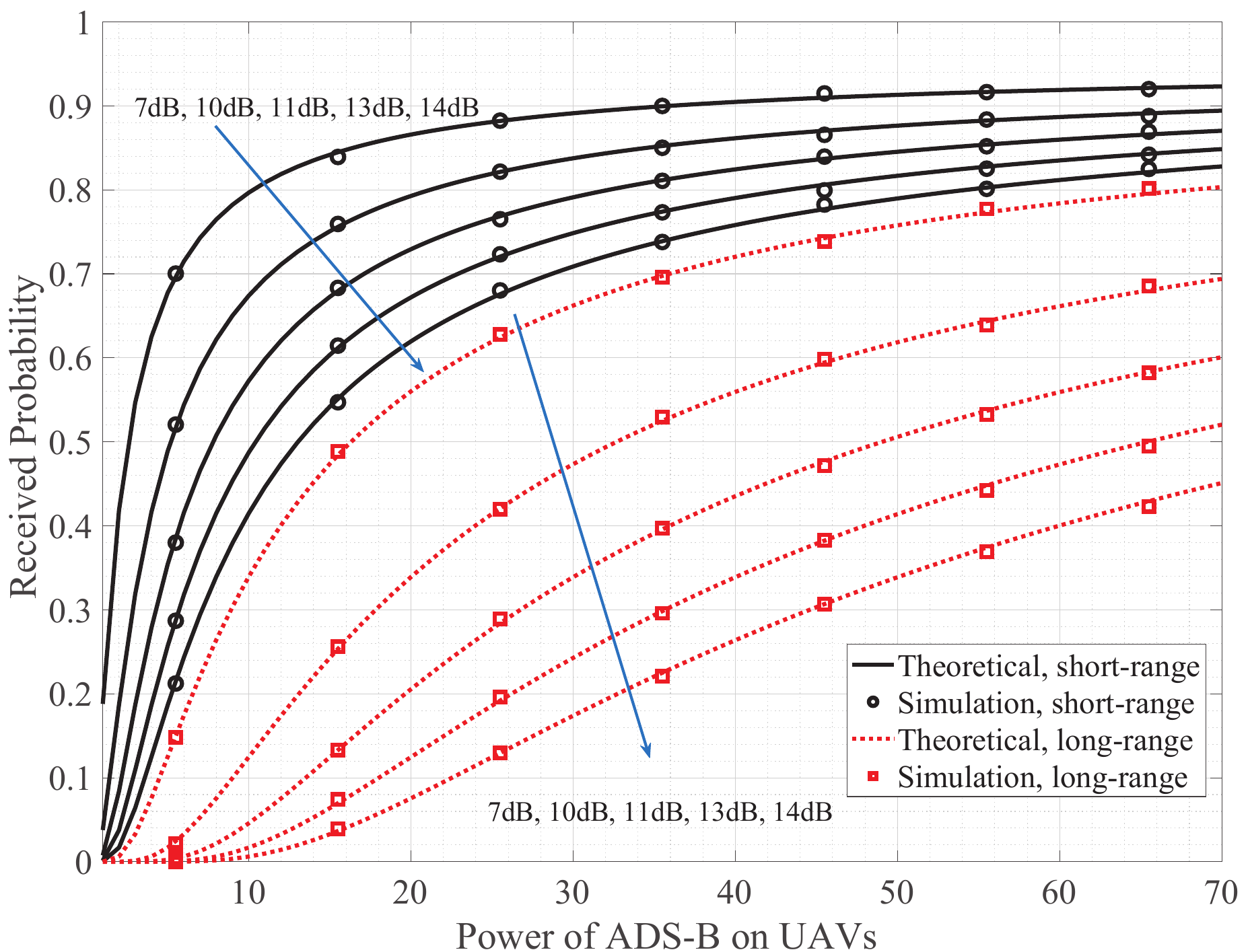}
 \label{f4}}
  \subfloat[The impact of $λ\lambda_1$ on the received probability.]{
 \includegraphics[width=0.31\linewidth]{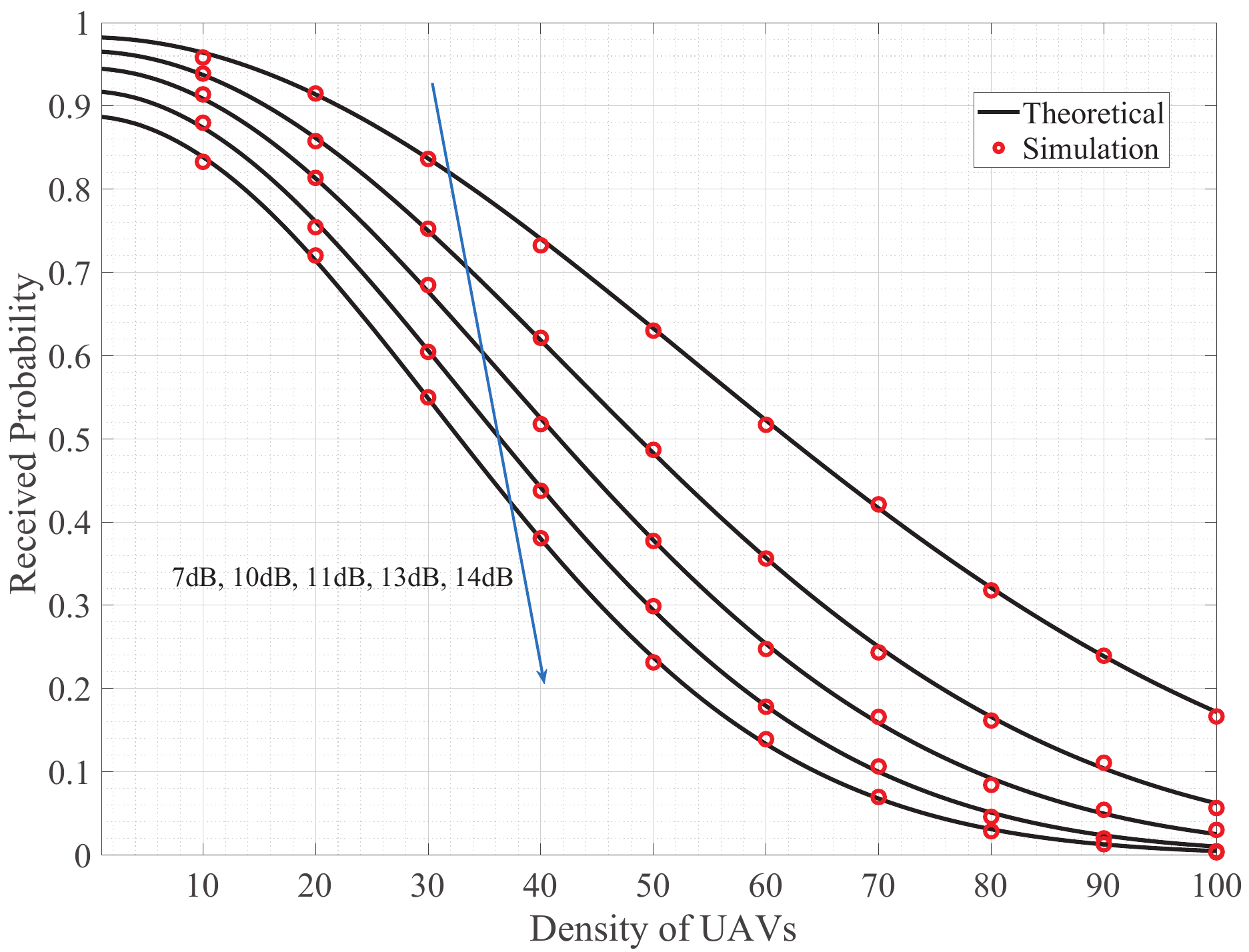}
  \label{f5}}
  \subfloat[The impact of $\alpha$ on the received probability.]{
  \includegraphics[width=0.31\linewidth]{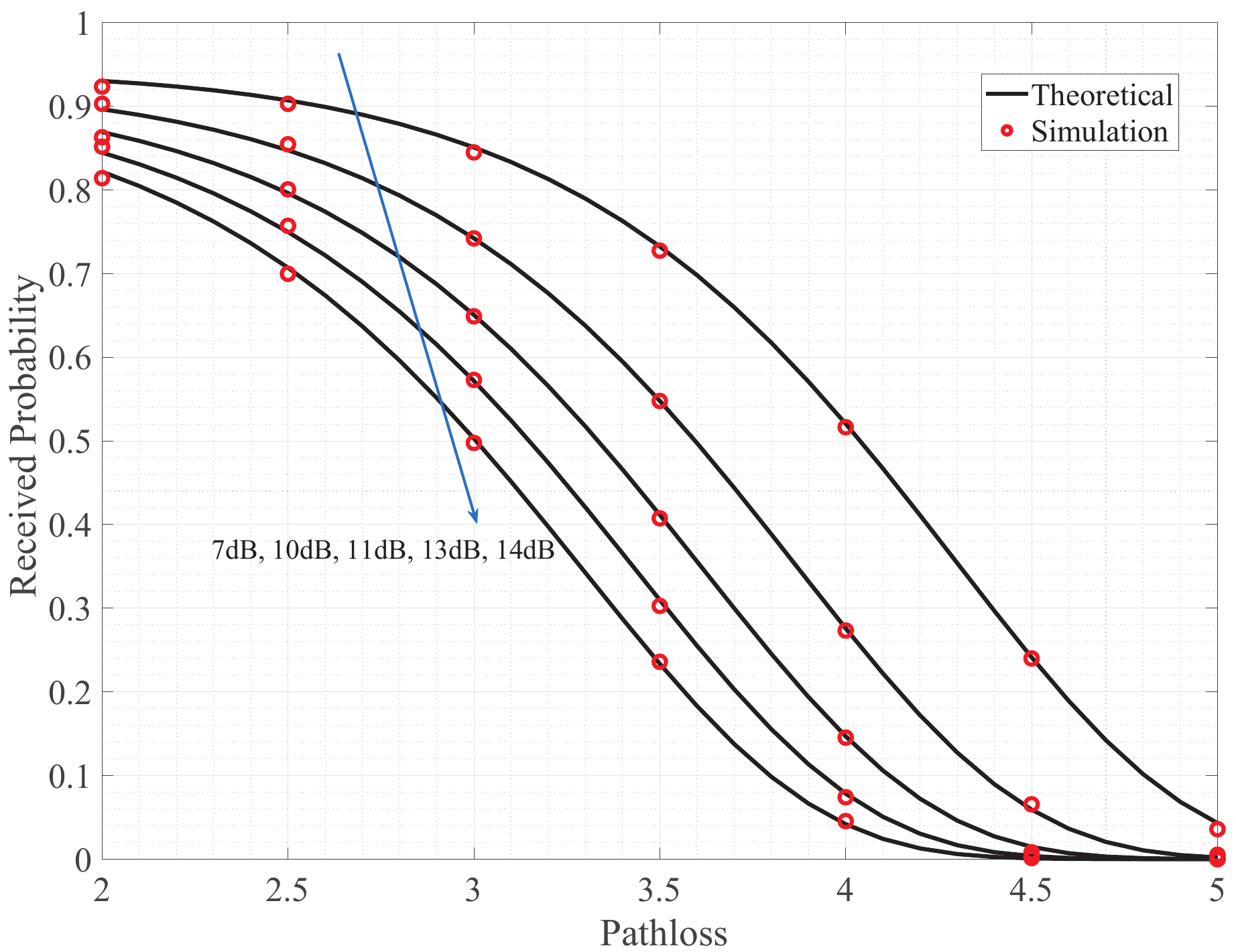}
 \label{f6}}
  \caption{Theoretical results and simulation results.}
  \label{f456}
  \end{minipage}
\end{figure*}

\begin{table}[t]
  \centering
  \caption{Key Parameters of the Simulation}
  \begin{center}
  \begin{tabular}{|p{1.0cm}<{\centering}|p{3.7cm}<{\centering}|}
  \hline  
  Parameter &Value \\
  \hline
  $B$&1MHz \\
  \hline
  $n_0$&-174dBm/Hz\\
  \hline
  $P_U$& 1W$\sim $70W\\
  \hline
  $G_U$&23dBi\\
  \hline
  $P_C$&15W$\sim$140W\\
  \hline
  $G_C$&20dBi\\
  \hline
  $\theta$& 7dB, 10dB, 11dB, 13dB, 14dB\\
  \hline
  $\alpha$& 2$\sim$5\\
  \hline
  $λ\lambda_1$& 0$\sim$100\\
  \hline
  $λ\lambda_2$& 15\\
  \hline
  $H_U$ & 1km$\sim$6km\\
  \hline
  $H_C$ & 6km$\sim$10km\\
  \hline
  \end{tabular}
  \label{tab1}
  \end{center}
  \end{table}

\section{Simulation Results and Analyses}\label{s4}
To further evaluate the detailed performance, MATLAB is employed to simulate the 3D-PPP distribution scenario of the UAVs and CAs, as shown in Fig. \ref{f2}. Space \textsl{\textbf{V}} is set as 20km$\times$ 20km$\times$10km. The CAs are randomly distributed at altitudes $H_C$ of $[6\rm km, 10\rm km]$, and the UAVs are randomly distributed at altitudes $H_U$ of $[1\rm km, 6\rm km]$ due to the limited flight ability. $λ\lambda_1$ is set within $[1, 100]$, and $λ\lambda_2$ is fixed as 15. In the civil aviation system, the channel interval within $[962\rm MHz, 1213\rm MHz]$ is 1MHz. Since we focus on ADS-B operating at 1090MHz, the channel bandwidth $B$ between the GS and all aircrafts are set as 1MHz\cite{ref16}. Besides, the pathloss index $\alpha$ is set within $[2, 5]$. Moreover, the total gain $G_U$ on AG channel of UAVs is 23dBi, and the total gain $G_C$ on AG channel of CAs is 20dBi. The detailed parameters in simulations are summarized in TABLE \ref{tab1}.

\par Fig. \ref{f3} shows the distance between the GS and UAVs or CAs. In detail, X-axis represents the generated aircrafts, including UAVs and CAs, and Y-axis denotes the distance between the aircrafts and the GS, corresponding to the simulation scenario in Fig. \ref{f2}. The distance of each point is randomly generated according to PPP. It is classified that if $d_{U_i}$$\textless$15km, $U_i$ is considered as a short-range UAV. Otherwise, if $d_{U_i}$$\geq$15km, $U_i$ is deemed as a long-range UAV.

  
    \par Fig. \ref{f4} demonstrates the impact of $P_U$ on the received probability under different threshold $\theta$, considering UAVs as targets. $P_C$, $\alpha$ and $λ\lambda_1$ are respectively fixed at 30W, 2 and 30. The solid lines represent the performance of short-range UAVs, while the dashed lines demonstrate the performance related to long-range UAVs. Besides, the simulation results coincide with the theoretical results, which validates the accuracy of theoretical analysis. As for the short-range UAVs, with the increment of $P_U$, the received probability from the desired UAVs at the GS increases. Moreover, as $\theta$ grows, the corresponding successful received probability decreases. When $\theta$ is 7dB and $P_U$ is 16W, the received probability is 84.77$\%$, where the received probability curve starts to smooth. If $\theta$ increases to 11dB, the received probability descends to 68.63$\%$. On the other hand, the dashed lines manifest the performance of the long-range flying UAVs. In particular,  $P_U$ has little influence on the received probability during the initial increment. Compared with short-range UAVs, with the increment of the distance, the received probability corresponding to the same $\theta$ and $P_U$ declines. When $\theta$ is 7dB and $P_U$ is 16W, the received probability is only 49.40$\%$, which decreases by 41.72$\%$ on the basis of short-range UAVs. If the threshold escalates, more transmitting power is needed to compensate for the received probability. In conclusion, for short-range UAVs, setting $P_U$ to be greater than 30W is energy-consuming, since the received probability is already approaching flat. Furthermore, in terms of long-range UAVs, the received probability improves significantly when $P_U$ falls between 30W and 70W\cite{ref17}. Additionally, the curve of $\theta$=7dB and $\theta$=10dB show smooth trends when $P_U$ exceeds 65W.

  \par To evaluate the density of UAVs, $λ\lambda_1$ in Fig. \ref{f5} is set within $[0, 100]$. We explore the impact of  $λ\lambda_1$ on received probability under different threshold $\theta$. $P_U$, $P_C$ and $\alpha$ are respectively fixed at 11dB, 30W and 2. The simulation results coincide with the theoretical results, which validates the accuracy of theoretical analysis. As the density of UAVs ascends, the received probability descends. Supposing $λ\lambda_1$ is 30, as we set in Fig. \ref{f4}, $\theta$=7dB, $\theta$=10dB, $\theta$=11dB, $\theta$=13dB, $\theta$=14dB correspond to the received probability of 83.68$\%$, 75.05$\%$, 67.69$\%$, 60.61$\%$, 54.86$\%$, respectively. Besides, when $\theta$ is higher 11dB and $λ\lambda_1$ exceeds 60, the GS can no longer support the monitoring of the airspace.

  \par Fig. \ref{f6} discusses the impact of pathloss $\alpha$ on the received probability under different threshold $\theta$. $P_U$, $P_C$ and $λ\lambda_1$ are respectively fixed at 25W, 30W and 30. The simulation results coincide with the theoretical results, which validates the accuracy of theoretical analysis. Primarily, $\alpha$ has little influence on the received probability during the initial increment. When $\alpha$ exceeds 3, the received probability drops sharply. Considering $\theta$ is 7dB and $\alpha$ is 3, the corresponding received probability is 85.1$\%$. When $\alpha$ rises to 4.5, the corresponding received probability declines to 24.1$\%$. If $\theta$ increases at this point, the GS can no longer support the monitoring of the airspace.
  In short, when $\alpha$ is greater 3, the increment cause the received probability to plummet, making the channel performance deteriorated.  By increasing $P_U$ or decreasing $\theta$, we can compensate for the effect of the increment of $\alpha$ on the received probability.


  \begin{figure}[t]
    \centerline{\includegraphics[width=0.88\linewidth]{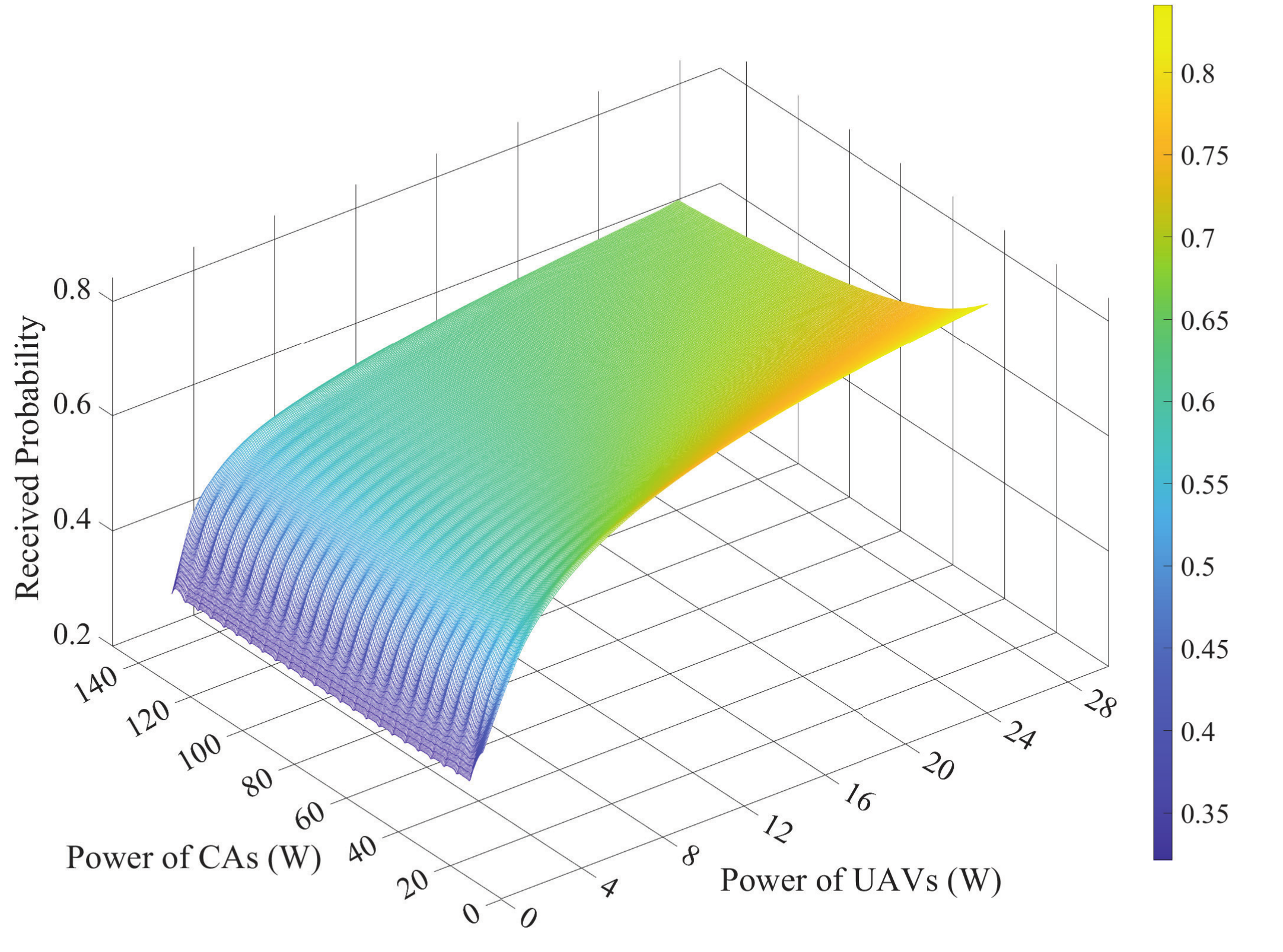}}
    \caption{The impact of $P_U$ and $P_C$ on the received probability.}
    \label{f7}
  \end{figure}  

\par Fig. \ref{f7} illustrates the impact of $P_U$ and $P_C$ on received probability. We set $P_C$ and $P_U$ as variables, aiming to simultaneously examines the influence of $P_U$ and $P_C$ on the received probability at short-range UAVs. $\theta$, $\alpha$ and $λ\lambda_1$ are  respectively fixed at 7dB, 2 and 30. With the increment of $P_U$, the received probability increases. On the contrary, as $P_C$ enlarges, the received probability diminishes. Assuming $P_U$ is 24W and $P_C$ is 40W, the corresponding received probability is 75.26$\%$. If $P_C$ stays constant, the corresponding received probability is 70.98$\%$ when $P_U$ decreases to 15W. If $P_U$ stays constant, the corresponding received probability is 70.49$\%$ when $P_C$ increases to 73W. Minishing $P_C$ improves the received probability of the GS towards UAVs. However, the monitoring performance of the GS towards CAs is undermined. In addition, $P_U$ is unadvisable to be magnified indefinitely, which intensifies the signal interference, impairs the ability of the GS to monitor CAs and wastes energy. Hence, the transmitting power of both types of aircrafts should be balanced according to actual demands.

\section{Conclusions}\label{s5}
This work analyzes the interference for the coexistence of UAVs and CAs based on ADS-B. We build a 3D-PPP model and deduces the explicit analytic form of the received probability targeting UAVs by SG theory. When analyzing the signal of a UAV, the interference signals are distinguished between UAVs and CAs. Moreover, based on the AG channel, we reveal the effects of transmitting power, density, threshold and pathloss on the performance of the ADS-B monitoring system via simulations. Additionally, the transmitting power of UAVs and CAs are both taken as variables to analyze the received probability of UAVs. In terms of raising the received probability, the transmitting power of ADS-B and the density of UAVs are contrarious, which can be set appropriately according to the requirements of the surveillance performance. In short, this work contributes to the appropriate deployment of ADS-B equipment on the UAVs, which helps improve the airspace safety and enhances the air traffic flow management.


\vfill
\end{CJK}
\end{document}